\documentclass[twocolumn,prl,superscriptaddress,showpacs,floatfix]{revtex4}

\usepackage{epsfig,color}
\usepackage{graphicx}
\usepackage{dcolumn}
\usepackage{bm}
\usepackage{setspace}

\begin{document}

\title{Temperature and Magnetic Field Enhanced Hall Slope
of a Dilute 2D Hole System in the Ballistic Regime}
\author{X. P. A. Gao}
\email{xuangao@lanl.gov}
\affiliation{Los Alamos National Laboratory, Los Alamos, NM 87545}
\author{G. S. Boebinger}
\affiliation{National High Magnetic Field Laboratory, Florida State University, Tallahassee, FL 32312}
\author{A. P. Mills Jr.}
\affiliation{Physics Department, University of California, Riverside, CA 92507}
\author{A. P. Ramirez, L. N. Pfeiffer, and K. W. West}
\affiliation{Bell Laboratories, Lucent Technologies, Murray Hill, NJ 07974}

\date{\today}

\begin{abstract}
We report the temperature($T$) and perpendicular magnetic field($B$)
dependence of the Hall resistivity $\rho_{xy}(B)$ of dilute metallic  
two-dimensional(2D) holes in GaAs over a broad range of temperature(0.02-1.25K). 
The low $B$ Hall coefficient, $R_H$, is found to be enhanced when $T$ decreases. 
Strong magnetic fields further enhance the slope of $\rho_{xy}(B)$ at all
temperatures studied. Coulomb interaction corrections of a 
Fermi liquid(FL) in the ballistic regime can not explain the enhancement of $\rho_{xy}$ 
which occurs in the same regime as the anomalous metallic longitudinal conductivity. 
In particular, although the metallic conductivity in 2D systems has been attributed to 
electron interactions in a FL, these same interactions should reduce, {\it not enhance}
the slope of $\rho_{xy}(B)$ as $T$ decreases and/or $B$ increases.   

\end{abstract}
\maketitle

The interplay between single particle localization and electron-electron 
interactions in
disordered electronic systems has been under much investigation for two 
decades\cite{Lee&rama}. Due to disorder induced single particle localization, 
2D non-interacting electron systems are predicted to be
insulators at zero temperature in the presence of any disorder\cite{Lee&rama}. 
It was also widely accepted that adding electron interactions does not
change this conclusion and, thus, there is no true metallic state in 2D at $T$=0. 
It came as a surprise when a 2D metallic state 
and metal-insulator transition(MIT) were observed in various high mobility low density 
2D systems after the initial discovery of Kravchenko {\it et al.}\cite{mitreview}.
The strong Coulomb interactions in these low density metallic systems revived  
interest in the role of Coulomb interactions in disordered 2D systems.  

A comprehensive theoretical understanding of the Coulomb interaction effects
on the 2D electron transport has emerged over the years\cite{Altshuler,Finkel,Gold,
DasSarma,Zala}. For diffusive electrons at low $T$, Coulomb interactions are known 
to give a ln$T$ conductivity correction $\delta\sigma(T)$, accompanying the similar 
ln$T$ correction from single particle interference in the weakly 
disordered regime\cite{Altshuler,Finkel}. 
Recently Zala, Narozhny and Aleiner(ZNA) pointed out that the 
logarithmic Altshuler-Aronov interaction correction to $\sigma$ originates from 
coherent scattering of Friedel oscillations. They extended the calculation to 
intermediate temperatures where transport is 
ballistic($k_BT>\hbar/\tau$) instead of diffusive($k_BT<\hbar/\tau$)\cite{Zala}. 
For high mobility samples exhibiting 2D metallic conduction, the elastic scattering
time $\tau$ is large and the sample is usually in the ballistic regime. 
In this regime, ZNA showed that $\delta\sigma(T)$, the 
interaction correction, could be positive('metallic')
or negative(insulating), depending on the FL parameter $F_0^\sigma$ just as in
the diffusive regime. The ZNA theory improves the previous screening theory
of Coulomb interactions at intermediate temperatures[5,6a,b], 
and predicts a linear $T$-dependent $\delta\sigma(T)$ controlled by $F_0^\sigma$.         
     
The interaction correction theory of FL systems in the ballistic regime\cite{Zala}
was applied by various experimental groups to explain the zero magnetic field 
metallic conductivity\cite{proskuryakov,coleridge,kvon,shashkin,Noh,pudalov,vitkalov}.
In these analyses, negative $F_0^\sigma$'s were obtained from 
fitting the metallic $\sigma(T)$ to a linear function of $T$ as predicted by the ZNA theory.
In the FL theory, a negative(positive) $F_0^\sigma$ corresponds to
ferromagnetic(antiferromagnetic) spin exchange interaction.  
While various scattering mechanisms besides the interaction correction can contribute
to the longitudinal conductivity, the $T$ dependent
Hall resistivity is a good probe for separating the Coulomb 
interaction effects\cite{Altshuler,bishophall,uren,emeleus}.
In this paper we present an analysis of the temperature 
dependent Hall resistivity together with the longitudinal conductivity of a 
metallic 2D hole system within the recent ballistic
FL theory in both weak\cite{Zala} and 
strong perpendicular magnetic field\cite{Gornyi}.     
We found that for all the densities studied, the slope of $\rho_{xy}(B)$ is 
enhanced by a decreasing temperature and/or increasing magnetic field. 
When the $B$=0 metallic conductivity is used to fix the FL parameters, analysis shows 
that the enhanced slope of $\rho_{xy}(B)$ is qualitatively and quantitatively inconsistent
with interaction corrections to Fermi liquid theory.
          
We performed the experiments on two dilute 2D hole systems in
two 10nm wide GaAs quantum wells.  The samples were made from the same 
wafer used in our previous study\cite{GaoPRL02}. The hole density
$p$ was tuned by a gold backgate which is about 150$\mu$m underneath the quantum well.
The two samples were measured in two different toploading Helium3-4 dilution refrigerators:
sample A was mounted on the copper tail of the mixing chamber of the refrigerator at 
UC-Riverside, while sample B was immersed in the liquid Helium3-4 mixture inside the mixing 
chamber of the refrigerator at LANL. The data collected from the two samples in the
two refrigerators are consistent with each other even down to our lowest experimental 
temperature of 20mK. During the measurements, the voltage applied to the sample was always kept low
(typically a few microvolts) such that the power delivered to the sample is less than a few
fWatts/cm$^2$ to avoid overheating the holes.

\begin{figure}[btph]
\centerline{\psfig{file=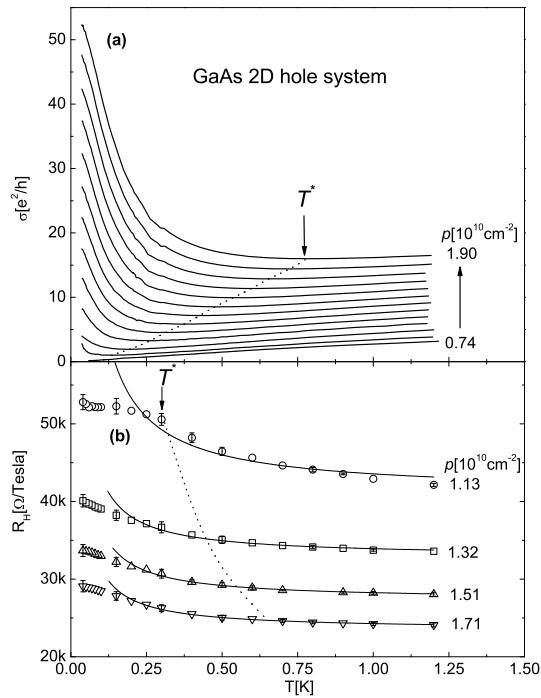,width=8cm}}
\caption{(a)The $B$=0 temperature dependent conductivity $\sigma(T)$ of 2D
holes with thirteen different densities in a 10nm wide GaAs quantum well (sample A).
The hole density spans from $p$=0.74 to 1.9$\times$10$^{10}$cm$^{-2}$ with 
0.965$\times$10$^{9}$cm$^{-2}$ step from the bottom curve to the top curve. 
The MIT of this sample happens around 
$p_c\sim$0.78$\times$10$^{10}$cm$^{-2}$.  
(b)The temperature dependent Hall coefficients for four densities in (a).
The black lines depict the functional behavior $const.+ 1/T$. For $p>p_c$, a dotted line is 
plotted in both panels to indicate the temperature $T^*$ where the sample turns metallic.}
\label{fig1}
\end{figure}

In Fig.\ref{fig1}a, we present the temperature dependent conductivity $\sigma(T)$ of 
sample A for various hole densities($p$=0.74-1.9$\times$10$^{10}$cm$^{-2}$) at $B$=0. The
density is determined from the Shubnikov-de Haas(SdH) oscillations. For all
the densities except 0.74$\times$10$^{10}$cm$^{-2}$, $\sigma(T)$ turns from 
insulating-like(d$\sigma(T)$/d$T>$0) to metallic-like(d$\sigma(T)$/d$T<$0) 
below a characteristic temperature $T^*$.
The metallic $\sigma(T)$ for $p>p_c$ below $T^*$ was recently attributed by some 
authors to the Coulomb interaction correction of a Fermi liquid with $F_0^\sigma<$0 
at intermediate temperatures according to the ZNA theory
\cite{proskuryakov,coleridge,kvon,shashkin,Noh,pudalov,vitkalov}.  
Theoretically, interaction effects will also give a correction to the Hall resistivity.
In the low $T$ diffusive limit, interactions have a correction $\delta R_H(T)\sim $ln$T$
to $R_H$, the Hall coefficient(the slope of $\rho_{xy}(B)$ in small $B$)\cite{Altshuler}.
In the ballistic regime, $\delta R_H(T)$ is expected to change to a 1/$T$ 
dependence\cite{Zala}. Thus, depending on the value of $F_0^\sigma$, 
$R_H$ will increase or decrease towards the Drude Hall coefficient as $R_H(T)\sim 1/T$ when 
$T$ increases.
Fig.\ref{fig1}b presents the $R_H$ vs. $T$ data for four metallic densities
in Fig.\ref{fig1}a. $R_H$ was obtained by linearly fitting $\rho_{xy}(B)$ between
-0.05T and +0.05T perpendicular field. 
It can be seen that at temperatures above 0.1K the measured $R_H(T)$ may be
described as a $const.+1/T$ function(Fig.\ref{fig1}b), although the fit fails at
lower temperatures where the theory should apply best.   

\begin{figure}[hbtp]
\centerline{\psfig{file=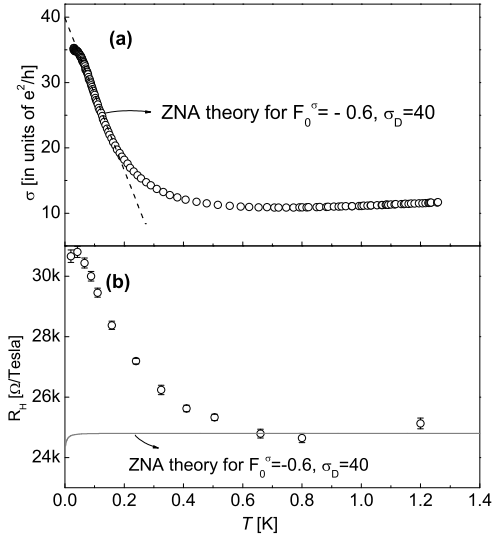,width=7.5cm}} 
\caption{(a) Conductivity $\sigma(T)$ for 2D holes 
with $p$=1.65$\times$10$^{10}$cm$^{-2}$ in sample B. 
The dashed black line is the linear fit of the $B$=0 metallic $\sigma(T)$ according
to the FL interaction correction theory of ZNA, which yields $F_0^\sigma$=-0.6
and Drude conductivity $\sigma_D$=40.
(b)Comparison of the $R_H(T)$ data for 2D holes in (a) with the theoretical 
expectation assuming the $B$=0 metallic conductivity is due to interaction correction 
of a FL. The gray line is the theoretical curve for $F_0^\sigma$=-0.6
and $\sigma_D$=40.
}
\label{fig2}
\end{figure}

Now we quantitatively discuss the longitudinal transport together with the Hall 
resistivity within the interaction correction theory of FL, using a density
($p$=1.65$\times$10$^{10}$cm$^{-2}$) in sample B as an example. 
Fig.\ref{fig2}a presents $\sigma(T)$ at $B$=0. 
In the ballistic regime, the interaction correction to conductivity is\cite{Zala}
\begin{equation}
\label{eq1}
\delta\sigma(T)=\sigma_D\left(1+\frac{3F_{\text{0}}^\sigma}{1+F_{\text{0}}^\sigma}\right)\frac{T}{T_F}.
\end{equation}
Following the analyses of ref.\cite{proskuryakov,coleridge,kvon,shashkin,Noh,pudalov,vitkalov}, 
we also can fit the $B$=0 conductivity data for 0.1K$<T<$0.2K to the linear dependence of 
Eq.~\ref{eq1}, obtaining a Drude conductivity of 40 $e^2/h$ and $F_0^\sigma$=-0.6.
The hole mass was set to be $m^*$=0.38$m_e$ in the fitting process, with
$m_e$ being the free electron mass.   
In Fig.\ref{fig2}b, $R_H$ vs $T$ data are plotted together with the predicted $R_H(T)$
(the gray line) according to ZNA theory with $\sigma_D$=40 $e^2/h$ and 
$F_0^\sigma$=-0.6. 
In the ZNA theory, the interaction correction to $R_H$ is the summation of 
the corrections from the singlet(charge) channel and the triplet(spin) channel:
$\delta R_H=\delta R^{\rho}_H+\delta R^{\sigma}_H$.
The singlet channel correction $\delta R^{\rho}_H$ and the triplet channel correction 
$\delta R^{\sigma}_H$ are given as Eq.17, Eq.18 respectively in ref.5c.

The discrepancy between the data and theoretical expectation in the metallic regime
of Fig.2 is obvious.
In fact, for $F_0^\sigma$=-0.6, the theory predicts a nearly flat but {\it decreasing} $R_H$
as temperature decreases in the experimental temperature range (20mK-1.2K).
Note that the FL theory predicts the interaction correction to $R_H$ to be 
very small in the ballistic regime for large $\sigma_D$, consistent with the Hall 
coefficient measurements for metallic 2D electrons in 
high mobility Silicon-metal-oxide-semiconductor field-effect transistors(Si-MOSFET's)
\cite{Pudalovhall,Sarachikhall,Khodas}.
   
\begin{figure}[htbp]
\centerline{\psfig{file=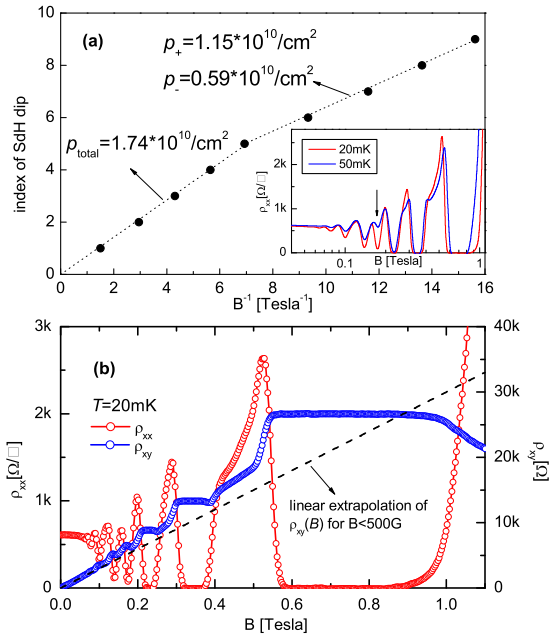,width=8.5cm}}
\caption{(color online)(a)The index number $i$ vs. $1/B$ of SdH dips for
sample B in Fig.2. 
Linear fitting $i$ vs. $1/B$ in high $B$ region yields a total hole density 
$p=1.74\times 10^{10}cm^{-2}$. Linear fitting the low $B$ part gives the densities 
for majority/minority spin subband $p_{+/-}=1.15,0.59\times 10^{10}cm^{-2}$. 
The inset shows the SdH oscillations at 20mK and 50mK, with an arrow marking
the beating node around 0.15T.
(b)Longitudinal resistivity $\rho_{xx}$ and Hall resistivity $\rho_{xy}$ at 20mK.
The quantized Hall plateaus and SdH minima coincide, yielding a density that is 
$\sim$20$\%$ smaller than deduced from the linear extrapolation (dashed line) of 
$\rho_{xy}$ from low fields($<$0.05T).}
\label{fig3}
\end{figure}
It is important to know if the temperature enhanced $R_H$ 
is actually related to a varying carrier density effect. 
A standard way to measure carrier density is the SdH oscillations in the longitudinal
magneto-resistivity $\rho_{xx}(B)$. From the positions of the SdH minima/maxima
one can extract the carrier density.   
At 20mK we could observe SdH oscillation 
in $\rho_{xx}(B)$ down to $\sim$0.06T.
Note that resolving SdH at low magnetic fields(high filling factors)
is difficult for low density holes with large effective mass(and hence small
cyclotron energy) because of the necessity to cool the holes to very low temperature. 
Fig.\ref{fig3}a shows the index number vs. 1/$B$ for the positions of the SdH oscillations
shown in the inset.  
We obtain the total hole
density $p=1.74\times 10^{10}cm^{-2}$ and the majority/minority spin subband
densities $p_{+/-}=1.15,0.59\times 10^{10}cm^{-2}$, via linear fitting of the index number
vs. 1/$B$ following ref.\cite{stormer,eisenstein}.  
The analysis of SdH beating is consistent with a fixed ($B$-independent) 
density(with 30$\%$ net spin polarization at $B$=0) 
in the regime of SdH oscillations and quantum Hall plateaus\cite{SdH}. However, 
the low-field($\leq$0.05T) slope of the Hall coefficient,$R_H(T)$, changes by 
more than 20$\%$ between 0.1 and 0.5K, temperatures sufficiently high that most SdH 
oscillations at high filling factors are no longer observable. Nevertheless, 
the positions of the SdH dips at $\nu$=1,2 do not move with $T$, and hence strongly 
imply a fixed ($T$-independent) carrier density. The $T$=20mK SdH oscillations and 
Hall resistivity $\rho_{xy}(B)$ are presented in Fig.\ref{fig3}b. 
The data are averaged from both
positive and negative magnetic field measurements to remove the admixture between
$\rho_{xx}$ and $\rho_{xy}$.  
We see that the 
SdH dips and quantized Hall plateaus occur at the same magnetic fields. 
Note, however, that the extrapolation of the low $B$($\leq$0.05T) $\rho_{xy}$ (dashed line)
intersects the Hall plateaus at magnetic fields higher than the plateau centers,
indicating that the low field $R_H$ is smaller than that determined at high fields.  
While this 20$\%$ discrepancy could, in principle, be due to 
interaction corrections to $R_H$\cite{Altshuler,bishophall,uren,emeleus}, 
we have already shown that the $\sigma(T)$ and $R_H(T)$ data are not explained 
consistently within the interaction theory of FL.

While ZNA's theory is only applicable in 
the low field limit ($\omega_c\tau<$1), Gornyi and Mirlin(GM) recently calculated the 
interaction correction to $\rho_{xy}$
into the high magnetic field regime($\omega_c\tau\gg$1) with $\omega_c=eB/m^*$ 
being the cyclotron frequency\cite{Gornyi}.  
We also investigated the behavior of $\rho_{xy}(B)$ in strong magnetic fields
to further test the FL interaction correction theory for our sample.
\begin{figure}[hbtp]
\centerline{\psfig{file=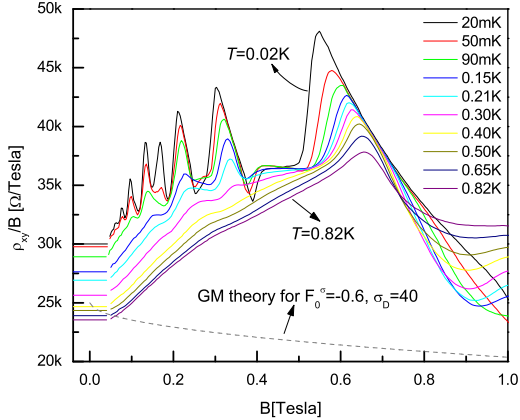,width=8cm}} 
\caption{(color online)
$\rho_{xy}$/$B$, the slope of Hall resistivity vs. magnetic field. 
When $B$ is strong($\omega_c\tau\gg 1$), a $\sqrt{B}$
dependent correction to $\rho_{xy}$/$B$ is expected in 
the FL theory\cite{Gornyi}. The dashed line is the theoretical curve for
parameters $F_0^\sigma$=-0.6 and $\sigma_D$=40 and a zero field value of
 $\rho_{xy}/B$=25k$\Omega$/T.  
}
\label{fig4}
\end{figure}
In GM's strong magnetic field theory, the interaction correction to $\rho_{xy}$
is separated into two parts. One part is $T$-dependent but $B$-independent, and 
the other part is $B$-dependent and $T$-independent. In Fig.\ref{fig4}, we plot the 
Hall slope, $\rho_{xy}/B$ vs. $B$ at various temperatures. To remove the admixture 
of $\rho_{xx}$ into $\rho_{xy}$, we antisymmetrized the $\rho_{xy}$ data
from both $B>$0 and $B<$0 measurements to obtain Fig.\ref{fig4} .   
The low field($B\leq$500G) $R_H$ data are also included.   
Fig.\ref{fig4} shows that the $\rho_{xy}/B$ data indeed may be viewed as a
$T$-independent magnetic field enhancement on the background of a $B$-independent
temperature enhancement\cite{notehall}.
The interaction correction to $\rho_{xy}$ at strong $B$ is also quantitatively
related to the FL parameter $F_0^\sigma$ as in ZNA theory\cite{Gornyi}. 
The $T$-dependent part of the $\rho_{xy}$ correction in the 
ballistic regime and strong $B$ is\cite{Gornyi}
\begin{equation}
\label{GMT}
\frac{\delta\rho_{xy}^T}{\rho_{xy}}=-7.117\frac{e^2/\hbar}{\sigma_D}\left(\frac{3F_0^\sigma}{F_0^\sigma+1}+1\right)\left(\frac{k_BT}{\hbar/\tau}\right)^{1/2}.
\end{equation}
In this high field regime, as in the low field regime, theory predicts a decreasing slope of 
the Hall resistivity with decreasing temperature. However, the opposite behaviour, i.e.
enhancement of the Hall resistivity, is observed when $T$ decreases. 
The $B$-dependent part of the 
GM correction to $\rho_{xy}$ is\cite{Gornyi} 
\begin{equation}
\label{GMB}
\frac{\delta\rho_{xy}^B}{\rho_{xy}}\approx\frac{e^2/\hbar}{\sigma_D}\left(\frac{3F_0^\sigma}{F_0^\sigma+1}+1\right)(\omega_c\tau)^{1/2}.
\end{equation}
Fig.\ref{fig4} also includes the theoretical curve from Eq.\ref{GMB} for 
$F_0^\sigma=-0.6$, $\sigma_D=40$ and $\rho_{xy}/B$($B$=0) = 25k$\Omega$/T.
One can see that $\delta\rho_{xy}^B/\rho_{xy}$ is expected to be negative for
$F_0^\sigma$=-0.6 but the data show a positive increase as $B$ increases.

Fig.\ref{fig4} also suggests that
$\rho_{xy}$/$B$ is enhanced with decreasing $T$ at both weak and strong magnetic 
fields in a similar fashion. It is reasonable to conclude that the $T$ dependent $\rho_{xy}/B$
originates from the same mechanism for both magnetic field regimes.
Since our temperature dependent SdH shows that the enhanced $\rho_{xy}/B$ at high $B$ is not 
related to a temperature dependent density, we further conclude that the enhanced low
magnetic field Hall coefficient is not due to a density effect. In conclusion, for
both the low magnetic field(ZNA) and high magnetic field(GM) regimes our combined resistivity
and Hall data are inconsistent with the electron interaction corrections interpretation in
a Fermi liquid.

Finally, we briefly comment on the relevance between our data and
several other FL-based models of the 2D metallic state, which do not invoke the 
FL parameters\cite{DasSarma,Altshulertrap}. 
For our sample in the metallic state, $\sigma$ is enhanced as large as three times as $T$ 
is reduced, a result perhaps consistent with the screening theory of Das Sarma and 
Hwang\cite{DasSarma}; however, the behavior of $R_H$ has not yet been 
theoretically discussed within the screening theory. Alternatively, the enhanced $R_H$ 
at low $T$ could be interpreted as a carrier freeze out[ref.6a] or trapping 
effect\cite{Altshulertrap};
however, the field($B>$0.06T) and temperature independent density we observe 
in the SdH oscillations require these effects to disappear above 0.06T and make these
interpretations seem highly unlikely.     
   
The authors are pleased to thank I.L. Aleiner and A. Punnoose for valuable discussions. 
Work at UCR was supported by LANL-CARE program. 
The NHMFL is supported by the NSF and the State of Florida.

\end{document}